\documentclass[
10pt,tightenlines, amsmath, amssymb, nofootinbib, prl,twocolumn,
superscriptaddress, showpacs, preprintnumbers]{revtex4}

\usepackage{graphicx}

\newcommand{\beq}{\begin{equation}}
\newcommand{\eeq}{\end{equation}}
\newcommand{\bea}{\begin{eqnarray}}
\newcommand{\eea}{\end{eqnarray}}
\newcommand{\nn}{\nonumber}

\newcommand{\junk}[1]{}

\begin{document}

 \title{ Confinement-Deconfinement Phase Transition
 at Nonzero Chemical Potential}
\author{D. Toublan}
\affiliation{Physics Department, University of Illinois at
  Urbana-Champaign, Urbana, IL 61801}
\author{Ariel R. Zhitnitsky}
\affiliation{Department of Physics and Astronomy, University of
  British Columbia, Vancouver, BC V6T 1Z1, Canada}
\date{\today}
\begin{abstract}
We present arguments suggesting that  large size
overlapping instantons 
are the driving mechanism of the confinement-deconfinement  phase
transition at nonzero chemical potential $\mu$. 
The arguments are based on the picture that instantons at  very large 
chemical potential in the weak coupling regime are localized  
configurations with finite size $\rho\sim\mu^{-1}$. At the same time, 
 the same instantons at  smaller chemical potential
  in the strong  coupling regime are well represented
 by the so-called instanton-quarks with fractional topological charge
 $1/N_c$.  
We  estimate the critical chemical potential
$\mu_c(T)$ where this phase transition takes place as a function of
temperature in the domain where our approach is justified.  
In this picture, the long standing problem of the ``accidental"
coincidence of the chiral  
and deconfinement phase transitions at nonzero temperature (observed
in lattice simulations) 
 is naturally resolved. We also derive results at nonzero
isospin chemical potential  $\mu_I$ 
 where direct lattice calculations are possible, and our predictions
 can be explicitly tested. 
 \end{abstract}
 
\maketitle

{\it   Introduction.} --- 
Color confinement, spontaneous breaking of chiral symmetry, 
the $U(1)$ problem and the $\theta$ dependence are some of the most
interesting questions in $QCD$. Unfortunately,  progress in the
understanding of these problems has been 
extremely slow. At the end of the 1970's A. M. Polyakov \cite{Po77}
demonstrated charge confinement in $QED_3$.  This was the first example
where nontrivial dynamics was shown to be a key ingredient for
confinement: 
The instantons  (the monopoles in 3d) play a crucial role
in  the dynamics of confinement in $QED_3$.  Soon afterwards
 instantons in four dimensional QCD were discovered
\cite{Belavin:1975fg}. However, their role in   $QCD_4$ remains
unclear due to the divergence  of the instanton density for 
large size instantons. 

Approximately at the same time instanton dynamics  was developed
in two dimensional, classically conformal, 
  asymptotically free models (which may have some analogies  with
  $QCD_4$). Namely, using an exact accounting and resummation
of the $n$-instanton solutions in $2d~CP^{N_c-1}$ models,
 the original problem of a statistical instanton
 ensemble  was   mapped unto a $2d$-Coulomb Gas (CG) 
system of pseudo-particles 
with fractional topological charges $\sim 1/N_c$ (the so-called
instanton-quarks) \cite{Fateev}.   The instanton-quarks do not exist
separately as individual objects. Rather, they appear in the system
all together as a set of $\sim N_c$ instanton-quarks so that the total
topological 
charge of each configuration is always an integer. This means that a
charge for an individual instanton-quark cannot be created
and measured.
Instead, only the total topological charge for the whole configuration
is forced to be integer and has a physical meaning. This picture leads to the 
elegant explanation of confinement and other important properties of
the $2d~CP^{N_c-1}$ models \cite{Fateev}.  
Unfortunately, despite some attempts \cite{Belavin}, there is no
demonstration that a
similar picture occurs in $4d$ gauge theories,  
where the instanton-quarks would become the relevant  quasiparticles. 
Nevertheless, there remains  a strong 
  suspicion that this picture, which assumes that instanton-quarks with
   fractional topological charges  $\sim 1/N_c$ become the relevant 
  degrees of freedom in the confined phase, may be correct in $QCD_4$.
 
 On the phenomenological side, the development of the 
instanton liquid model (ILM) \cite{ILM, shuryak_rev} has encountered 
  successes (chiral symmetry breaking, resolution of the $U(1)$
 problem, etc) and 
 failures (confinement could not be described by well separated and
 localized lumps with integer 
 topological charges). Therefore, it is fair to say that
 at present, the widely accepted viewpoint is that
 the  ILM can explain  many experimental data
  (such as hadron masses, widths, correlation functions, decay
  couplings, etc), with one, but crucial exception: confinement.
  There are many arguments against the ILM approach, see e.g. \cite{Horvath}, 
  there are many arguments supporting it \cite{shuryak_rev} .

  In this letter we present  new arguments supporting the idea
  that the instanton-quarks 
  are the relevant quasiparticles in the strong coupling regime. In this case, 
  many problems
formulated in \cite{Horvath} are naturally resolved as both phenomena,
confinement and chiral symmetry breaking are originated from the same
vacuum configurations, instantons,  
which may have arbitrary scales: the finite sized localized lumps, as well as 
 set of $N_c$ fractionally  charged  $1/N_c$ correlated objects with
 arbitrary large separations.  
 
More importantly, 
  we make some very specific predictions which can be tested with 
  traditional Monte Carlo techniques, by studying QCD at nonzero isospin
  chemical potential\cite{Isospin}.  
  We start by reviewing  recent work for QCD 
at large $\mu$ in the deconfined phase \cite{ssz}, where the
  instanton calculations are under complete  
  theoretical control, since the instantons are
  well-localized objects with a typical size 
  $\rho\sim 1/\mu$.
We then discuss the dual representation of the
low-energy effective chiral Lagrangian in the regime
of small chemical potential where confinement takes place.
 We shall argue that the corresponding dual representation
 corresponds to a statistical 
system of interacting pseudo-particles with fractional $1/N_c$ topological
charges  which can be  identified   with instanton-quarks \cite{JZ} 
suspected long ago \cite{Fateev,Belavin}.
Based on these observations  we   {\bf conjecture} that the transition from 
the description in terms of well localized instantons with finite size
at large $\mu$  to the description 
in terms of the instanton quarks with fractional $1/N_c$ topological
charges precisely corresponds to the deconfinement-confinement phase
transition.
In what follows we explicitly calculate the critical chemical
potential $\mu_c$ where this phase transition occurs. Our conjecture
can be explicitly and readily tested in numerical simulations
due to the absence  of the  sign problem  at arbitrary value of the 
isospin chemical potential. If our conjecture turns out to be
correct, it would be   
an explicit  demonstration of the link between confinement and instantons.
 
 {\it Instantons at large $\mu$.---}
At low energy and large chemical potential, the $\eta'$ is light and 
described by the Lagrangian derived in \cite{ssz}:
\begin{eqnarray}
  \label{Leta}
  L_\varphi=f^2(\mu) [(\partial_0\varphi)^2-u^2 (\partial_i\varphi)^2] -
  V_{inst}(\varphi). 
\end{eqnarray}
where the $\varphi$ decay constant, $f^2(\mu_B)\!=\!\mu_B^2/8\pi^2$ and
$f^2(\mu_I)\!=\!3\mu_I^2/16\pi^2$, and its velocity,
$u^2\!=\!1/3$ \cite{ssz,schaefer}. 
We define baryon and isospin chemical potentials as
$\mu_{B,I}\!=\!(\mu_u\pm\mu_d)/2$.    
The nonperturbative potential $V_{inst}\sim \cos(\varphi-\theta)$
is due to instantons, which are suppressed at large chemical potential.
  
The instanton-induced effective four-fermion interaction for 2
flavors, $u, d$, is given by
\cite{tHooft,SVZ},
\begin{eqnarray}
   L_{\rm inst}&=& \int\!d\rho\, n(\rho) 
  \biggl({4\over3}\pi^2\rho^3\biggr)^2 \biggl\{
  (\bar u_R u_L)(\bar d_R d_L) + \nonumber \\
  &+& {3\over32} \biggl[ (\bar u_R\lambda^a u_L)(\bar d_R\lambda^a d_L)
\nonumber\\
  &-& {3\over4}(\bar u_R\sigma_{\mu\nu}\lambda^a u_L)
    (\bar d_R\sigma_{\mu\nu}\lambda^a d_L) \biggr]
  \biggr\}  + {\rm H.c.}
  \label{inst_vertex}
\end{eqnarray}
We study this problem at nonzero temperature and chemical
potential for $T\ll\mu$, and we use the standard formula
for the instanton density at two-loop order \cite{shuryak_rev}
\begin{eqnarray}
 n(\rho)&=& C_N(\beta_I(\rho))^{2N_c} \rho^{-5}
 \exp[-\beta_{II}(\rho)]  \\
 && \hspace{.5cm}\exp[-(N_f \mu^2 + \frac13 (2N_c+N_f) \pi^2
 T^2)\rho^2], \nn 
\end{eqnarray}
where
\begin{eqnarray}
  C_N &=& 0.466 e^{-1.679N_c} 1.34^{N_f}/(N_c-1)!(N_c-2)! \nn\\
\beta_I(\rho)&=&-b \log(\rho\Lambda), \;\;
\beta_{II}(\rho)=\beta_I(\rho)+\frac{b'}{2b} \log\left(\frac{2
  \beta_I(\rho)}{b}\right), \nn \\ 
b&=& \frac{11}3 N_c-\frac23 N_f, \;\;
 b'=\frac{34}3 N_c^2-\frac{13}3 N_f
N_c +\frac{N_f}{N_c}.\nn
\end{eqnarray}
By taking the average of Eq.\ (\ref{inst_vertex}) over the
 state  with nonzero vacuum expectation value for the
 condensate,  one finds  
\bea
  V_{\rm inst}(\varphi)&=& -\int\!d\rho\, n(\rho)
  \biggl({4\over3}\pi^2\rho^3\biggr)^2 12|X(\mu)|^2\cos(\varphi-\theta)~~~
  \nn \\ 
 &=& -a(\mu,T)\mu^2\Delta^2\cos(\varphi-\theta),
\label{Vinst2}
\eea 
where  $|X(\mu_B)|\!=\!3\mu_B^2\Delta/\sqrt{\beta_I(\rho)}$ and
$|X(\mu_I)|\!=\!3\sqrt{3}\mu_I^2\Delta/\sqrt{\beta_I(\rho)}$, and
$\Delta$ is the gap \cite{ssz,schaefer}.  
Therefore the mass of the $\varphi$ field is given by
\begin{eqnarray}
  \label{massEta}
  m=\sqrt{\frac{a(\mu,T)}2} \frac{\mu \Delta}{f(\mu)}.
\end{eqnarray}
The approach presented above is valid as long as the $\varphi$ field is
lighter than $\sim 2\Delta$, the mass of the other mesons in the
system \cite{ssz}, that is  if
\begin{eqnarray}
  \label{critical}
  a(\mu,T) \lesssim 8 f^2(\mu)/\mu^2. 
\end{eqnarray}
This is exactly the vicinity where the Debye screening scale and the
inverse gap become of the same order of magnitude \cite{ssz}, and
therefore, where the instanton expansion breaks down.

For reasons which will be clear soon, we want to represent 
 the  Sine-Gordon (SG) partition function (\ref{Leta}, \ref{Vinst2}) in
the equivalent dual Coulomb Gas (CG) representation \cite{ssz},
\bea
\label{CG}
Z = \sum_{M_\pm=0}^\infty \frac{(\lambda/2)^M}{M_+!M_-!}  
\int d^4x_1 \ldots \int d^4x_M ~
e^{-i\theta\sum_{a=0}^M Q_a}\cdot  \nn\\
e^{-{1\over 2f^2u}\sum_{a>b=0}^M  Q_aQ_b
G(x_a-x_b)} ,~~~~~~~~~~~~~\nn\\
G(x_a - x_b) = {1\over 4\pi^2 (x_a-x_b)^2}, ~~
\lambda \equiv {a\mu^2\Delta^2\over u}. ~~
\eea
Physical interpretation of the dual CG representation (\ref{CG}):\\
a) Since
$Q_{\rm net}\equiv \sum_a Q_a$ is the total charge and it appears in
the action
multiplied be the parameter $\theta$, one
concludes that $Q_{\rm net}$ {\it is} the total topological
charge  of a given
configuration.\\
b) Each charge 
$Q_a$ in a 
given configuration should be identified with an
  integer topological charge   well localized at the point $x_a$. This, 
by definition, 
corresponds to a small instanton positioned at $x_a$.\\
c) While the starting low-energy effective Lagrangian contains only 
a   colorless field $\varphi$  we
have ended up with a representation of the partition function in which 
  objects carrying color (the instantons)   can be studied.\\
d)
In particular, $II$ and $I\bar I$
interactions (at very  large distances) are exactly the same up to a
sign, order $g^0$, 
and are Coulomb-like.  This is in  contrast with semiclassical expressions
when  $II$ interaction is zero and $I\bar I$  interaction is order $1/g^2$.\\
e) The very complicated picture of the    bare  $II$ and $I\bar I$
interactions 
becomes very simple for  dressed   instantons/anti-instantons
when all integrations over all  possible sizes, color orientations
and interactions with background fields are properly accounted for.\\
f) As expected, the ensemble of small $\rho\sim 1/\mu$ instantons can
not produce confinement. 
 This is in accordance with the fact that there is no confinement at
 large $\mu$. 
 
{\it Instantons at small $\mu$}~\cite{JZ}. ---
 We   want to repeat the same procedure that led to the CG representation
 in the confined phase at small $\mu$ to see if any traces from the
 instantons  
 can be recovered. We start from the chiral Lagrangian and keep only
 the diagonal   
 elements of the chiral  matrix 
 $U=\exp\{i{\rm
diag}(\phi_1,\dots,\phi_{N_f})\}$ which are relevant in the
description of the ground state. 
 Singlet  combination is defined as  $\phi =
{\rm Tr}~U$.
  The effective Lagrangian for  the $\phi$  is
 \bea
 \label{chiral}
 L_{\eta'}= f^2 
( \partial_{\mu} \phi)^2 + E \cos\left( \frac{ \phi - \theta }{N_c}\right) +
\sum_{a=1}^{N_f} m_a \cos \phi_a
\eea
  A    Sine-Gordon structure for the singlet combination  
  corresponds to the following behavior of the $(2k)^{\rm
th}$ derivative of the vacuum energy in pure gluodynamics \cite{veneziano}.
$$ \left.
\frac{ \partial^{2k} E_{vac}(\theta)}{ \partial \, \theta^{2k}}
\right|_{\theta=0} \sim \int \prod_{i=1}^{2k} dx_i \langle
Q(x_1)...Q(x_{2k})\rangle \sim (\frac{i}{N_c})^{2k},
$$ 
where $Q=\frac{g^2}{32\pi^2} G_{\mu\nu} {\widetilde G}_{\mu\nu}$ is
the topological density.  
The same structure was also advocated in \cite{HZ} from a different
perspective. 
 As in (\ref{CG}) the Sine-Gordon effective field
 theory (\ref{chiral}) can be represented in terms 
 of a classical statistical ensemble (CG representation) given by
 (\ref{CG})  with the replacements $\lambda\rightarrow E, ~ u\rightarrow 1$. 
 The fundamental difference in comparison with the previous case
 (\ref{CG}) is that
while the total charge is integer, the individual charges are
   { \bf fractional $\pm 1/N_c$}.
This  is a
direct   consequence of the $\theta/N_c$ dependence in the underlying
effective Lagrangian (\ref{chiral}) before integrating out $\phi$ fields.\\
 Physical Interpretation of the  CG representation (\ref{CG}) of
 theory (\ref{chiral}):\\  
 a) As before,  one can identify
$Q_{\rm net}\equiv \sum_a Q_a$ with  the   total topological charge of
the given configuration.\\ 
b) Due to the $2\pi$
periodicity of the theory, only configurations which contain an integer
topological number contribute to the partition function. Therefore, 
the number of particles for each given configuration $Q_i$ with
charges $ \sim 1/N_c$  
 must be proportional to $N_c$.\\
c) Therefore, the number of integrations
over $d^4x_i$   in CS representation exactly equals $4 N_c k$, where 
$k$ is integer.   This number  $4 N_c k$ exactly
corresponds to the number of zero modes in the  $k$-instanton
background. This is basis for the conjecture \cite{JZ}  that 
at low energies (large distances) the fractionally charged species,
$Q_i=\pm 1/N_c$   
are the {\bf   instanton-quarks} suspected long ago \cite{Fateev}.\\
d) For  the gauge group, $G$
the number of integrations would be equal to  $4k C_2(G)$  where
$C_2(G)$ is the quadratic Casimir of the gauge group
  ($\theta$ dependence in physical observables comes  in
the combination $\frac{\theta}{C_2(G)}$).   This number  $4k C_2(G)$
exactly corresponds 
to the  number of zero modes  in the $k$-instanton background for gauge
group $G$.\\
e) The CG representation  corresponding to eq.(\ref{chiral}) describes
the confinement  phase of the theory.
One immediate objection: 
  it has long been known  that   instantons 
can explain most low energy QCD phenomenology \cite{ILM}   with the
 exception   confinement;  and we claim
that   confinement arises in this picture: how can this be consistent?   
 We note that quark confinement can not be described in the  dilute gas 
approximation, when the instantons and anti-instantons are well
separated and maintain their individual properties (sizes, positions,
orientations), as at large $\mu$.  However,   
in strongly coupled theories the instantons and
anti-instantons lose their individual properties (instantons will
``melt'')  their sizes become very large and they overlap.  The relevant
description is that of instanton-quarks  which can be far away from
each other,  but still  strongly correlated. 
 
{\it Conjecture.} ---
We thus conjecture that the confinement-deconfinement phase transition
takes place at precisely the value where the dilute instanton
calculation breaks down:  At low $\mu$ 
  color is confined (because of the
instanton-quarks), whereas at large $\mu$ 
 color is not confined (because of dilute instantons).  The value of
the critical chemical potential as a function of temperature,
$\mu_c(T)$ is given by saturating the inequality (\ref{critical}).

Few remarks are in order. We can calculate the temperature dependence
of $\mu_c(T)$ 
only at relatively low $T$ where our calculations are justified.
We also note that the critical chemical potentials 
given below  are not sensitive to the
specific assumptions made in the derivation of eq. (\ref{chiral}) 
as the numerical estimates below are based on approaching critical values from 
  large $\mu$. We expect that our numerical results for $\mu_c(T)$ are
  not very sensitive to   
  many unavoidable uncertainties   due to the large
  power of $(\Lambda_{QCD}/\mu)^b$  entering the instanton density.
 
 {\it Results.} ---
 The critical chemical potential as a function of temperature is
 implicitly given by $a(\mu_c(T),T)=8 f^2(\mu_c(T))/\mu_c(T)^2$.  
We can calculate $a(\mu_c(T),T)$ from (\ref{Vinst2}).  We
 are however limited to temperatures where Cooper pairing takes place,
i.e. for $T\lesssim0.567 \Delta$ \cite{pr}.
We have determined the critical chemical potential in different cases
  at nonzero baryon or isospin chemical potential.  We
find that the values of
the critical chemical potentials at $T\!=\!0$ are 
given by (we use $m_s\!\simeq\!150$~MeV  which is numerically close
to $ 0.75\Lambda$ for $N_f\!=\!3$)\footnote{
If we   used 1-loop $\beta$ function, we  would get
larger coefficients in the table. It is  simple  reflection of the
fact that we are very close to the critical values where the
interaction  is essential. While numerical predictions are not robust,
  the  general picture of the transition  is not sensitive to the
  details and  remains the same.}: 

\vspace{.2cm}
\begin{tabular}[c]{|c|c|c|c|}
\hline 
 & $N_c\!=\!3,\; N_f\!=\!2$ & $N_c\!=\!3,\; N_f\!=\!3$ & $N_c\!=\!2,\;
 N_f\!=\!2$ \\ 
\hline 
$\mu_{Bc}/\Lambda$ & 2.3 & 1.4 & 3.5 \\
\hline
$\mu_{Ic}/\Lambda$ & 2.6 & 1.5 & 3.5 \\
\hline
\end{tabular}
\vspace{.2cm}

\begin{figure}[h]
\hspace{-.8cm}
\includegraphics[scale=0.3, clip=true, angle=0,
draft=false]{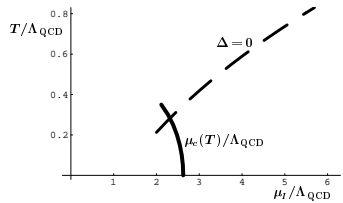}
\caption{\label{fig1} Critical isospin chemical potential for the
  confinement-deconfinement phase transition as a
  function of temperature (solid curve). The dashed curve represents
  the largest temperatures that can be reached in our approach, 
  given by $0.567 \Delta$ (see text for more details).}
\end{figure}

As an example, we explicitly show the
results as a function of temperature 
for $N_c=3$ at nonzero $\mu_I$ in FIG.~\ref{fig1}, 
where direct lattice calculation are possible. We notice that with our
conventions the transition from the normal phase to pion condensation
happens at $\mu_I\!=\! m_{\pi}/2$. 

\begin{figure}[h]
\hspace{-0.8cm}
\includegraphics[scale=0.25, clip=true, angle=0,
draft=false]{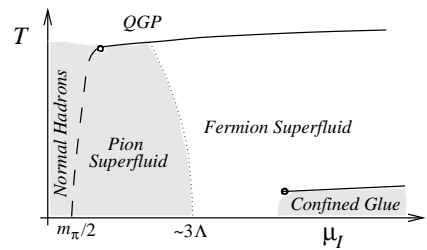}
\caption{\label{fig2} Phase diagram of QCD at nonzero temperature and
  isospin chemical potential.  First and second order 
phase transitions are depicted by solid and dashed curves,
respectively. Chiral
symmetry is broken everywhere except in the QGP phase. 
The confined phases are shaded.  
Confinement of pure glue is expected at very  low $T$ and 
very large $\mu_I$ \cite{Isospin}.} 
\end{figure}

{\it Conclusion.} --- 
In this article we have conjectured that there is a
confinement-deconfinement phase transition at nonzero chemical
potential and small temperature that is driven by instantons (which
transformed  
from well-localized objects to strongly overlapped configurations).
Furthermore we make a
quantitative prediction for the critical value of the chemical
potential where this transition takes place: $\mu_c\!\sim\!3\Lambda_{\rm
  QCD}$ at $T\!=\!0$.  This prediction can be
readily tested on the lattice at nonzero isospin chemical potential.
Our conjecture corresponds to the statement that the
confinement-deconfinement transition and the topological charge
density distribution (instanton-quark to instanton transition)
must experience  
sharp changes exactly at the same critical value $\mu_c(T)$. 
There are well- established lattice methods which allow to measure
the topological charge density distribution, see e.g.
\cite{Gattringer, Horvath}.
Independently, there are well established lattice method which allow to 
introduce isospin chemical potential into the system, see
e.g. \cite{kogut}. 
We claim that the topological charge density distribution measured as
a function  of $\mu_I$ 
will experience sharp changes at the same critical value
$\mu_I=\mu_{c}(T)$  
where the phase transition (or rapid crossover) occurs.
If our conjecture is correct, the phase diagram of QCD 
at nonzero temperature
and isospin chemical potential should be given by
FIG.~\ref{fig2}. 

Finally, at the intuitive level there seems
 to be a close relation between our conjecture about   
instanton quarks and the ``periodic instanton" analysis
\cite{vanbaal, diakonov, Gattringer}. 
Indeed, in these papers  it has been shown that the large size
instantons and monopoles are intimately connected
and  instantons have the internal structure resembling the
instanton-quarks.  
Unfortunately,
one should not expect to be able to account
for large instantons using semiclassical technique   
to bring this intuitive correspondence onto the quantitative level.
However, this analogy may help us to  understand   the relation 
between picture advocated by   't~Hooft and Mandelstam \cite{Hooft}
and confinement due to the 
instanton-quarks, as conjectured in the present paper. 
The key
point of the 't~Hooft - Mandelstam approach is the assumption that
dynamical monopoles exist and Bose condense. 
If our conjecture is correct, then one can argue
(on the basis of semiclassical analysis \cite{vanbaal}, see also
\cite{JZ})  that 
the instanton-quarks carry the magnetic charges and are responsible
for confinement. In this case both
pictures could be the two sides of the same coin. 

The authors thank Pierre van Baal for discussions. We also thank   the KITP 
for the organization of the workshop
"Modern Challenges for Lattice Field Theory"  
 which initiated and motivated this study  in the course of
 discussions on the subject \cite{inst_quark}.
 The work of
A.R.Z. was supported, in part, by the Natural Sciences and Engineering
Research Council of Canada.  D.T. is supported in part by NSF grant
NSF-PHY-0102409.


\begin{thebibliography}{1234567}
\bibitem{Po77}
A. M. Polyakov, Nucl. Phys. {\bf B120} (1977) 429.
\bibitem{Belavin:1975fg}
  A.~A.~Belavin, A.~M.~Polyakov, A.~S.~Shvarts and Y.~S.~Tyupkin,
  Phys.\ Lett.\ B {\bf 59}, 85 (1975).
\bibitem{Fateev} V.Fateev et al, Nucl. Phys. {\bf B154} 
(1979) 1;
B.Berg and M.Luscher, Commun.Math.Phys. {\bf 69}(1979) 57.
\bibitem{Belavin} A. Belavin et al, Phys. Lett. {\bf 83B} (1979) 317.
\bibitem{ILM} 
  E.~V.~Shuryak,
  Nucl.\ Phys.\ B {\bf 203}, 93 (1982);
  D.~Diakonov and V.~Y.~Petrov,
  Nucl.\ Phys.\ B {\bf 245}, 259 (1984);
 \bibitem{shuryak_rev} T.~Schafer and E.~V.~Shuryak,
  Rev.\ Mod.\ Phys.\  {\bf 70}, 323 (1998).
\bibitem{Horvath}
  I.~Horvath, N.~Isgur, J.~McCune and H.~B.~Thacker,
  Phys.\ Rev.\ D {\bf 65}, 014502 (2002);
I.~Horvath {\it et al.},
  Phys.\ Rev.\ D {\bf 66}, 034501 (2002)

\bibitem{Isospin}
  D.~T.~Son and M.~A.~Stephanov,
  Phys.\ Rev.\ Lett.\  {\bf 86}, 592 (2001);
  Phys.\ Atom.\ Nucl.\  {\bf 64}, 834 (2001).


\bibitem{ssz}
D.~T.~Son, M.~A.~Stephanov, and A.~R.~Zhitnitsky, 
Phys.Rev.Lett. {\bf 86}, 3955 (2001);
Phys. Lett. {\bf B510}, 167 (2001).


\bibitem{JZ} S. Jaimungal and A.R. Zhitnitsky,hep-ph/9904377,hep-ph/9905540,
unpublished.


\bibitem{schaefer}
T.~Schafer,
  Phys.\ Rev.\ D {\bf 67}, 074502 (2003);
  Phys.\ Rev.\ D {\bf 65}, 094033 (2002).

\bibitem{tHooft}
G.~'t Hooft,
Phys.\ Rev.\  D {\bf 14}, 3432 (1976).

\bibitem{SVZ}
M.A.~Shifman, A.I.~Vainshtein, and V.I.~Zakharov,
Nucl.\ Phys.\  {\bf B163}, 46 (1980);
{\bf B165}, 45 (1980).

\bibitem{veneziano} G. Veneziano,  Nucl. Phys. {\bf B159} 
(1979) 213.
 \bibitem{HZ} I. Halperin and A.R.  Zhitnitsky,{\it Phys. Rev. Lett.} 
{\bf81}, 4071 (1998); {\it Phys. Rev.} {\bf D58}, 054016 (1998).
  
\bibitem{pr}  
R.~D.~Pisarski and D.~H.~Rischke,
  Phys.\ Rev.\ D {\bf 61}, 051501 (2000).
  \bibitem{Gattringer}
  C.~Gattringer,
  Phys.\ Rev.\ D {\bf 67}, 034507 (2003);
  C.~Gattringer and S.~Schaefer,
  Nucl.\ Phys.\ B {\bf 654}, 30 (2003).

\bibitem{kogut} 
  J.~B.~Kogut and D.~K.~Sinclair,
  Phys.\ Rev.\ D {\bf 66}, 034505 (2002);
  Phys.\ Rev.\ D {\bf 70}, 094501 (2004).


\bibitem{vanbaal} T.C. Kraan and P. van Baal,
Nucl. Phys. {\bf B533} (1998) 627; Phys. Lett.{\bf  B428}  (1998) 268,
Phys. Lett.{\bf  B435}  (1998) 389;
M.G.Perez, T.G.Kovacs, P. van Baal Phys. Lett.{\bf  B472}  (2000) 295,
hep-ph/0006155;
\bibitem{diakonov}
  D.~Diakonov, N.~Gromov, V.~Petrov and S.~Slizovskiy,
  Phys.\ Rev.\ D {\bf 70}, 036003 (2004);
  D.~Diakonov and N.~Gromov,
  arXiv:hep-th/0502132.
\bibitem{Hooft} G't Hooft, in "Recent Developments in Gauge Theories"
Charges 1979, Plenum Press, NY 1980;
Nucl. Phys. {\bf B190} (1981) 455; 
S. Mandelstam, {\it Phys. Rep.} {\bf 23} (1976) 245. 

\bibitem{inst_quark} A. Zhitnitsky, Instanton Quarks,  talk at KITP
http://online.kitp.ucsb.edu/online/lattice05/zhitnitsky/
\end{thebibliography}
\end{document}